\def\Re{{\rm Re}}
\documentclass [prb,twocolumn,superscriptaddress,showpacs]{revtex4}
\usepackage {bm}
\usepackage {graphicx}

\begin{document}

\title{
X-ray fluorescence holography: going beyond the diffraction limit
}

\begin{abstract}
X-ray fluorescence holography (XFH) is a method for obtaining 
diffraction-limited images of the local 
atomic structure around a given type of emitter. The reconstructed 
wave-field represents a distorted 
image of the scatterer electron density distribution; {\it i.e.} it is a 
convolution of the charge density distribution 
with a point spread function characteristic of the measurement. We here 
consider several methods for the 
iterative deconvolution of such XFH holograms, and via theoretical 
simulations evaluate them from the point 
of view of going beyond the diffraction limit so as to image the 
electron charge density. Promising results for 
future applications are found for certain methods, and other possible 
image-enhancement techniques are also discussed.
\end{abstract}

\author{S. Marchesini}
\affiliation{Materials Sciences Division, Lawrence Berkeley National Laboratory,
Berkeley, California 94720}
\author{C. S. Fadley}
\affiliation{Materials Sciences Division, Lawrence Berkeley National Laboratory, Berkeley, California 94720}
\affiliation{Department of Physics, University of California, Davis, California 95616}
\author {F. J. Garcia de Abajo}
\affiliation{Centro Mixto, CSIC-UPV/EHU, San Sebastian, Spain}
\noaffiliation


\date{\today}
\maketitle

\section{Introduction}

X-ray fluorescence holography (XFH) is a promising method for directly 
imaging local atomic structure in an element-specific way that has been 
developed experimentally over the past six years 
\cite{AIP:1986,Tegze:1996,Gog:1996}. In the first mode of 
measurement, an atom inside an oriented system ({\it e.g.} a single crystal with 
low to medium mosaicity) emits a fluorescent x-ray wave; the interference 
between the outgoing unperturbed reference wave component and the scattered 
object wave component produces a hologram 
\cite{AIP:1986,Tegze:1996}. One can term this ``inside 
source'' holography. A time-reversed mode also exists, in which the exciting 
x-ray wave is the reference, and scattered x-rays impinging on a fluorescent 
emitter are the object waves \cite{Gog:1996}. The emitter is the 
detector of the hologram, and thus this can be called ``inner detector'' 
holography. The holographic reconstruction in either case is achieved by 
propagating back the wavefield from the far field.

Beyond several recent achievements of these methods 
\cite{Fitzgerald:2000}, {\it e.g.} in imaging the environment around a dilute 
semiconductor dopant \cite{Hayashi:1998}, low-Z atoms in the presence of 
intermediate-Z atoms \cite{Tegze:2000}, and the local environment in a 
quasicrystal \cite{Marchesini:2000}, another step toward the practical 
application of holography would be a quantitative link between the 
electronic charge distribution and the holographic image. It has already 
been demonstrated that images with accuracies at the diffraction limit can 
be obtained \cite{Tegze:1999}. But even in this case, the reconstructed 
wavefield can be viewed as a distorted image of the scatterer density 
distribution; {\it i.e.} it is the convolution of the charge density distribution 
with a point spread function due to the measurement and inversion process.

When an object is imaged through an optical system, information is lost 
whenever the imaging system cannot pass all the spatial frequencies 
contained in the scene. Further loss is produced by aberrations in the 
optical system. The question that we here address is how much information 
can be recovered if we know the point spread function (PSF) of the effective 
optical system associated with x-ray fluorescence holography. We will 
consider here several methods for the iterative deconvolution of data so as 
to improve image quality.

Several procedures for iterative image improvement have been developed 
previously in the fields of astronomy and microscopy \cite{Stark:1987}, 
and we begin by introducing them briefly. Van Cittert developed a very early 
method in 1920 \cite{Van:1970}; this uses a self-consistent iterative 
solution to the problem. It can be shown that this method is close to the 
gradient search of the maximum likelihood (ML). Two other approaches are 
based on the probability theory. One developed by Lucy \cite{Lucy:1974} 
and Richardson \cite{Richardson:1972} is based on Bayes theorem for 
conditional probability. Another one, the maximum entropy method (MEM) seeks 
the `most probable' solution in an under-determined system of equations 
\cite{Press:1992}. In this article, we discuss some of these methods that 
seem most appropriate for applying to XFH image improvement, quantitatively 
assess several of them via model theoretical calculations, and finally 
arrive at the first deconvolved atomic image at resolution going beyond the 
diffraction limit.

\section{Basic imaging concepts in XFH}

We first consider some basic imaging concepts for XFH, in particular for an 
arbitrary atomic charge density distribution $\rho \left( {\rm {\bf r}} 
\right)$. For simplicity, we will also throughout this discussion consider 
only ``inner source'' holography, although the conclusions here can easily be 
generalized to include ``inner detector'' holography. Neglecting absorption or 
extinction effects (a reasonable approximation if we consider only 
near-neighbor imaging), the hologram $\chi $ in the far field $\rm {\bf k}$ is 
given by \cite{Faigel:1999}:
\begin{eqnarray}
\nonumber
 \chi \left( {\rm {\bf k}} \right) &=& \int {\eta \left( {{\rm {\bf k}},{\rm 
{\bf r}}} \right)\rho \left( {\rm {\bf r}} \right)\,d^3{\rm {\bf r}}} 
\;,\mbox{ with} \\ 
 \eta \left( {{\rm {\bf k}},{\rm {\bf r}}} \right) &=& 2\Re\left[ 
{\textstyle{{r_{e} } \over r}e^{i(kr - {\rm {\bf k}} \cdot {\rm {\bf 
r}})}f\left( {{\rm {\bf k}},{\rm {\bf r}}} \right)} \right]\,;  
\label{eq1}
 \end{eqnarray}
\noindent
here $r_{e} $ is the classical electron radius, $r_{e} f\left( {{\rm {\bf 
k}},{\rm {\bf r}}} \right)$ is the scattering factor per electron and the 
phase factor is given by the path length difference between the wave emitted 
from the origin and the wave scattered by the electron located in ${\rm {\bf 
r}}$. The typical reconstruction algorithm to calculate the wavefield $U$ in 
the real space ${\rm {\bf r}}$ is \cite{Barton:1988}:
\begin{eqnarray}
\nonumber
 U\left( {\rm {\bf {{r}'}}} \right) &=& \frac{1}{\Omega }\int_\sigma {\eta 
^{\{ - 1\}}\left( {{\rm {\bf {{r}'}}},{\rm {\bf k}}} \right)\chi \left( {\rm 
{\bf k}} \right)\,d^3{\rm {\bf k}}} ,\mbox{ with} \\ 
 \eta ^{\{ - 1\}}\left( {{\rm {\bf {{r}'}}},{\rm {\bf k}}} \right) &=& 
\textstyle{{{r}'} \over {r_{e} }}\Re\left[ {e^{ - i\left( {k{{r}'} - {\rm 
{\bf k}} \cdot {\rm {\bf {{r}'}}}} \right)}} \right]; 
\label{eq2}
 \end{eqnarray}
\noindent
here $\sigma $ is the measured region in k-space, $\Omega $ the volume of 
$\sigma$, and we have included the scaling factor 
$\frac{{{r}'}}{r_{e} }$ to obtain an estimate of the charge density.  
Sometimes an additional factor $f^{\{ - 1\}}$ is included to compensate the 
`aberration' or distortions produced in the hologram $\chi $ by 
$f$\cite{Tonner:1991}, but we will not explicitly include this here. We 
will for simplicity assume that the scattering factor of an electron is 
isotropic, neglecting the angular dependence of the Thompson scattering and 
near field effects, The holographic reconstruction U can thus be considered 
to be a distorted image of the charge density distribution $\rho $:
\begin{eqnarray}
\nonumber
 U\left( {{\rm {\bf r}}'} \right) &=& \int {\mu \left( {{{\rm {\bf r}}'},
\rm {\bf r}} 
\right)\rho \left( {\rm {\bf r}} \right)\,d^3 {\rm {\bf r}}} , \mbox{ with}\\ 
 \mu \left( {{\rm {\bf {{r}'}}},{\rm {\bf r}}} \right) &=& 
\int_\sigma {d^3{\rm {\bf k}}\;\eta ^{\{ - 1\}}\left( {{\rm {\bf 
{{r}'}}},{\rm {\bf k}}} \right)\eta \left( {{\rm {\bf k}},{\rm {\bf r}}} 
\right)} ;  
\label{eq3}
 \end{eqnarray}
\noindent
with $\mu \left( {{\rm {\bf {{r}'}}}\mbox{,}{\rm {\bf r}}} \right)$ now 
being the experimental point spread function. An additional factor $\left( {1 + 
{\rm sign} U\left( {{\rm {\bf {{r}'}}}\mbox{,}{\rm {\bf r}}} \right)} \right) / 2$ 
can be used to enforce positivity. In summary we have three spaces ${\rm 
{\bf r}}$, ${\rm {\bf k}}$ and ${\rm {\bf {{r}'}}}$ for which we have three 
functions $\rho \left( {\rm {\bf r}} \right)$, $\chi \left( {\rm {\bf k}} 
\right)$ and $U({\rm {\bf {{r}'}}})$ and the propagators from one space to 
the other are $\eta \left( {{\rm {\bf k}},{\rm {\bf r}}} \right)$, $\eta 
^{\{ - 1\}}\left( {{\rm {\bf k}},{\rm {\bf {{r}'}}}} \right)$ and $\mu 
\left( {{\rm {\bf {{r}'}}}\mbox{,}{\rm {\bf r}}} \right)$.

We now turn to specific methods for iterative deconvolution of images. For 
notational brevity, we in the following sections use the symbol $ \cdot $ 
between two functions to denote integration over the internal variable, 
{\it e.g.} $\quad f(x,y) \cdot g(y) = \int {dy\,f(x,y)g(y)} $.

\section{Iterative image deconvolution}
\label{sec:iterative}
The general problem of the deconvolution is to obtain the unknown object 
$\rho \left( {\rm {\bf r}} \right)$ from a knowledge of $U({\rm {\bf 
{{r}'}}})$ and $\mu ({\rm {\bf {{r}'}}},{\rm {\bf r}})$. The first method 
developed by Van Cittert in 1920 \cite{Van:1970} is the most 
intuitive one and it is based on a self-consistent solution. Every step is 
the difference between the reconstructed experimental hologram $U\left( {\rm 
{\bf r}} \right)$ and the reconstructed simulated hologram $U^{\{n\}}\left( 
{\rm {\bf r}} \right) = \mu \left( {{\rm {\bf r}},{\rm {\bf {{r}'}}}} 
\right)\ast \rho ^{\{n\}}\left( {\rm {\bf r}} \right)$ generated by a new 
$n^{th}$-step charge density distribution. That is,
\begin{eqnarray}
\nonumber
 \rho ^{\left\{ {n + 1} \right\}}\left( {\rm {\bf r}} \right) &=& \rho 
^{\left\{ n \right\}}\left( {\rm {\bf r}} \right) + \Delta \rho ^{\left\{ n 
\right\}}\left( {\rm {\bf r}} \right), \\ 
\nonumber
 \Delta \rho ^{\left\{ n \right\}}\left( {\rm {\bf r}} \right) &=& \left[ 
{U\left( {\rm {\bf r}} \right) - U^{\left\{ n \right\}}\left( {\rm {\bf r}} 
\right)\,} \right], \\ 
 \rho ^{\left\{ 1 \right\}}\left( {\rm {\bf r}} \right) &=& U\left( {\rm {\bf 
r}} \right)\,. 
\label{eq4}
 \end{eqnarray}
Thus, $\mu ({\rm {\bf {{r}'}}},{\rm {\bf r}})$ is in this approximation 
assumed to be a delta function $\delta ({\rm {\bf {{r}'}}},{\rm {\bf r}})$ 
over each step in the iteration. We normally enforce positivity after each 
step by setting to 0 the negative values of the charge density.

A similar problem is the phase retrieval of a diffraction pattern of known 
intensity from an object with known support $s\left( {\rm {\bf r}} \right)$. 
The diffraction pattern provides the amplitude of Fourier transform of the 
charge density $\left| \tilde {\rho } \right|$. If the object $\rho \left( 
{\rm {\bf r}} \right)$ is constrained within a given support ({\it e.g.} with the 
support equal to 1 over the object and 0 outside of it), the product of the 
object $\rho \left( {\rm {\bf r}} \right)$ with the support function 
$s\left( {\rm {\bf r}} \right)$ must be still equal to the object. Therefore 
the complex amplitude of the diffracted wave-field is not modified by a 
convolution with the Fourier transform of the support function $\tilde {s}$: 
\begin{equation}
\label{eq5}
\rho = s\rho \buildrel {FT} \over \longrightarrow \tilde {\rho } = \tilde 
{s}\ast \tilde {\rho }\,.
\end{equation}
If $\tilde {s}$ is sufficiently large (`over-sampling' condition 
\cite{Miao:1999}) a certain number of pixels are bounded by this 
self-consistent equation. The main difference with iterative deconvolution 
methods is that in this case, the bigger $\tilde {s}$ is, the larger is the 
number of pixels bounded by (\ref{eq5}), and the easier it is to find the 
solution. If the over-sampling condition is satisfied, iterative approaches 
such as the Gerchberg-Saxton \cite{Gerchberg:1972} and Fienup 
\cite{Fienup:1982} algorithms can be used to solve the phase problem. In 
the Fourier domain, the Gerchberg-Saxton algorithm can be expressed as:
\[
\tilde {\rho }^{\left\{ {n + 1} \right\}} = \left| {\tilde {\rho 
}_{\mbox{meas}} } \right|e^{i{\kern 1pt} \mbox{phase}\left\{ {\tilde {\rho 
}^{\left\{ n \right\}}\ast \tilde {s}} \right\}},
\]
\noindent
where the $n^{th}$ and $(n+1)^{th}$ estimates for density are given, and $\left| 
{\tilde {\rho }_{\mbox{meas}} } \right|$ is the measured amplitude of the 
Fourier transform of the object. 

Another iterative approach is given by the Lucy-Richardson method 
\cite{Lucy:1974,Richardson:1972}. This method is based on a 
self-consistent iterative solution of the Bayes theorem for conditional 
probabilities, considering $\mu \left( {{\rm {\bf {{{r}'}}}},{\rm {\bf 
{r}}}} \right)$ as a probability that ${\rm {\bf {{r}'}}}$ will fall in the 
interval $\left( {{\rm {\bf {{{r}'}}}},{\rm {\bf {{{r}'}}}} + d{\rm {\bf 
{{{r}'}}}}} \right)$ when it is known that ${\rm {\bf r}} = {\rm {\bf 
{{{r}'}}}}$. A self-consistent iterative approach can in this case be 
expressed as \cite{Lucy:1974}:
\begin{eqnarray}
\nonumber
 \rho ^{\left\{ {n + 1} \right\}}\left( {\rm {\bf r}} \right) = &\rho 
^{\left\{ n \right\}}\left( {\rm {\bf r}} \right) \\ 
&\times \left[ {\mu ^T\left( {{\rm {\bf r}},{\rm {\bf {{r}'}}}} \right)\ast 
\frac{U\left( {\rm {\bf {{r}'}}} \right)}{\mu \left( {{\rm {\bf 
{{r}'}}},{\rm {\bf {r}''}}} \right)\ast \rho ^{\left\{ n \right\}}\left( 
{\rm {\bf {r}''}} \right)}} \right] 
\label{eq6}
\end{eqnarray}
\noindent
with $\mu ^T({\rm {\bf r}},{\rm {\bf {{r}'}}})$ being the transpose of $\mu 
$.
\section{Maximum likelihood and entropy}
\label{sec:maximum}
In another set of methods, we want to minimize the difference between the 
measured hologram in the far field $\chi \left( {\rm {\bf k}} \right)$ and 
the hologram generated by a given charge density distribution $\rho 
^{\left\{ n \right\}}\left( {\rm {\bf r}} \right)$:
\begin{equation}
\label{eq7}
{\rm {\bf Q}} = \int_\sigma {d^3{\rm {\bf k}}\,\left| {\chi \left( {\rm {\bf 
k}} \right) - \eta \left( {{\rm {\bf k}},{\rm {\bf r}}} \right) \cdot \rho 
^{\left\{ n \right\}}\left( {\rm {\bf r}} \right)} \right|} ^2\,.
\end{equation}
\noindent
where the unknowns are $\rho ^{\left\{ n \right\}}\left( {{\rm {\bf r}}_{i} } 
\right),\;i = 1...N$ over some grid spanning the object. The gradient can be 
computed as:
\begin{equation}
\label{eq8}
\nabla {\rm {\bf Q}} = 2\eta ^\dag \left( {{\rm {\bf r}},{\rm {\bf k}}} 
\right) \cdot \left[ {\eta \left( {{\rm {\bf k}},{\rm {\bf r}}} \right) 
\cdot \rho ^{\left\{ n \right\}}\left( {\rm {\bf r}} \right) - \chi \left( 
{\rm {\bf r}} \right)} \right]\,.
\end{equation}
\noindent
with $\eta ^\dag $ being the transpose conjugate of $\eta $. The Hessian of 
${\rm {\bf Q}}$ is in this case nothing else but the point spread function 
and is given by:
\begin{eqnarray}
\nonumber
 {\rm {\bf H}} &=& \frac{\partial ^2{\rm {\bf Q}}}{\partial \rho ^{\left\{ n 
\right\}}\left( {{\rm {\bf r}}_{1} } \right)\partial \rho ^{\left\{ n 
\right\}}\left( {{\rm {\bf r}}_{2} } \right)} \\ 
 &=& 2\eta ^\dag \left( {{\rm {\bf r}}_{1} ,{\rm {\bf k}}} \right) \cdot \eta 
\left( {{\rm {\bf k}},{\rm {\bf r}}_{2} } \right). 
\label{eq9}
 \end{eqnarray}
The single pixel approximation \cite{Wilkins:1983,Cornwell:1985} 
consists in assuming that ${\rm {\bf H}}$ is diagonal, so that the steps 
will be:
\begin{equation}
\label{eq10}
\Delta \rho ^{\left\{ n \right\}}\left( {\rm {\bf r}} \right) = - 
\frac{1}{{\rm {\bf H}}_{{\rm {\bf r}},{\rm {\bf r}}} }\nabla {\rm {\bf 
Q}}_{{\rm {\bf r}}} \;.
\end{equation}
Under this approximation, the gradient search is equivalent to the Van 
Cittert method, since $H_{{\rm {\bf r}},{\rm {\bf r}}} \simeq \left( 
{\textstyle{{r_{e} } \over r}} \right)^2$ (see proof in Appendix 
\ref{sec:appendix}) and $\eta ^{\{ - 1\}} = \textstyle{{r^2} \over 
{r_{e}^{2} }}\eta ^\dag $. However, for single energy holograms, at the 
inversion-symmetric position to the real atoms, there will appear a twin or 
ghost image. Therefore one should consider at least the symmetric term ${\rm 
{\bf H}}_{{\rm {\bf r}}, - {\rm {\bf r}}} $:
\begin{eqnarray}
\label{eq11}
{\rm {\bf \bar {H}}}_{{\rm {\bf r}},{\rm {\bf {{r}'}}}} &=& {\rm {\bf 
H}}_{{\rm {\bf r}},{\rm {\bf {{r}'}}}} \left( {\delta _{{\rm {\bf 
{{r}'}}},{\rm {\bf r}}} + \delta _{{\rm {\bf {{r}'}}}, - {\rm {\bf r}}} } 
\right)\\
\label{eq12}
\Delta \rho ^{\left\{ n \right\}}\left( {\rm {\bf r}} \right) &=& - \left( 
{{\begin{array}{*{20}c}
 {{\rm {\bf \bar {H}}}_{{\rm {\bf r}},{\rm {\bf r}}} } \hfill & {{\rm {\bf 
\bar {H}}}_{{\rm {\bf r}}, - {\rm {\bf r}}} } \hfill \\
 {{\rm {\bf \bar {H}}}_{ - {\rm {\bf r}},{\rm {\bf r}}} } \hfill & {{\rm 
{\bf \bar {H}}}_{ - {\rm {\bf r}}, - {\rm {\bf r}}} } \hfill \\
\end{array} }} \right)^{ - 1}\left( {{\begin{array}{*{20}c}
 {\nabla {\rm {\bf Q}}_{{\rm {\bf r}}} } \hfill \\
 {\nabla {\rm {\bf Q}}_{ - {\rm {\bf r}}} } \hfill \\
\end{array} }} \right).
\end{eqnarray}
An additional consideration here however, is that at every energy, for a 
given position, and in case of a point scattering, the twin image cancels 
the real image, so that a divergence arises in Eq. (\ref{eq12}). This can be 
avoided by introducing an anisotropy in the scattering factor between the 
forward and backward directions. It is known \cite{Szoke:1993} that, if 
we use a small Gaussian charge distribution for every point ${\rm {\bf 
r}}_{i} $ instead of point scatterers, the scattering becomes stronger in 
the forward direction, and iterative algorithms converge more quickly. 
Conjugate gradient methods do not require the calculation of the Hessian, 
but partial knowledge of it is helpful. Extra information can also be used 
to speed up the process, and for example, we know that the scattering charge 
distribution is real and positive. This can be enforced by taking the real 
part at every step and by setting to $\rho _{n} = 0$ when $\rho _{n} \left( 
{\rm {\bf r}} \right) < 0$. A further piece of information is the total 
number of scattering electrons ${\rm {\bf F}}_{0} $, and we can use Lagrange 
multipliers to perform a constrained best fit via:
\begin{eqnarray}
\nonumber
 {\rm {\bf Q}} + \beta \,\left( {{\rm {\bf F}} - {\rm {\bf F}}_{0} } 
\right),\mbox{ with} \\ 
 {\rm {\bf F}} = \sum {\rho \left( {{\rm {\bf r}}_{i} } \right)} .
\label{eq13}
 \end{eqnarray}
Here the minimization is performed adjusting the multiplier  at every 
step:
\begin{eqnarray}
\nonumber
 \Delta \rho _{{\rm {\bf r}}} &=& - {\rm {\bf \bar {H}}}_{{\rm {\bf r}},{\rm {\bf 
{{r}'}}}} ^{ - 1}\left( {\nabla {\rm {\bf Q}}_{{\rm {\bf {{r}'}}}} + \beta 
\;\nabla {\rm {\bf F}}_{{\rm {\bf {{r}'}}}} } \right), \\ 
 \lambda &=& \frac{\left[ {{\rm {\bf F}}_{0} - {\rm {\bf F}} + \left( {\nabla 
{\rm {\bf F}} \cdot {\rm {\bf \bar {H}}}^{ - 1} \cdot \nabla {\rm {\bf Q}}} 
\right)} \right]}{\nabla {\rm {\bf F}} \cdot {\rm {\bf \bar {H}}}^{ - 1} 
\cdot \nabla {\rm {\bf F}}} 
\label{eq14}
 \end{eqnarray}
The Lagrange multiplier method keeps the number of electron fixed, however 
after we apply each step, we enforce positivity in the charge density by 
setting to 0 the negative values of the charge density, therefore the actual 
number of electrons will tend to grow at every iteration. In the actual 
implementation of the algorithm, we artificially reduce the step size by a 
factor of 1/5 to avoid cutting too much  in the negative density regions.

The deconvolution of spectra with an incomplete Fourier image is also often 
treated with the maximum entropy method \cite{Press:1992}. When the 
number of unknowns $\rho \left( {{\rm {\bf r}}_{i} } \right)$ is superior to 
the number of equations, it is necessary to include further \textit{a-posteriori} information. As 
more than one solution can satisfy the set of equations, we seek the most 
probable solution, {\it i.e.} the one with the maximum entropy. In practice we 
want to minimize the functional:
\begin{eqnarray}
\nonumber
 &{\rm {\bf S}}& + {{\lambda }'}{\rm {\bf Q}}, \\
 &{\rm {\bf S}}& = \sum\nolimits_{i} {\rho \left( {{\rm {\bf r}}_{i} } \right)\ln 
\rho \left( {{\rm {\bf r}}_{i} } \right)} 
\label{eq15}
 \end{eqnarray}
\noindent
where ${{\lambda }'}$ is a Lagrange multiplier. The procedure is similar to 
the previous one, with two main differences: (i) MEM enforces positivity, so 
that a pure gradient minimization could lead to unphysical values that need 
to be chopped, (ii) ${\rm {\bf S}}$ is highly nonlinear, and simple gradient 
search is not efficient. An interesting property of ${\rm {\bf S}}$ is that 
its Hessian is diagonal; therefore the single pixel approximation can be 
used \cite{Cornwell:1985}.

\section{Numerical simulations}
\label{sec:numerical}
\begin{figure}[htbp]
\centerline{\includegraphics[width=3.5in]{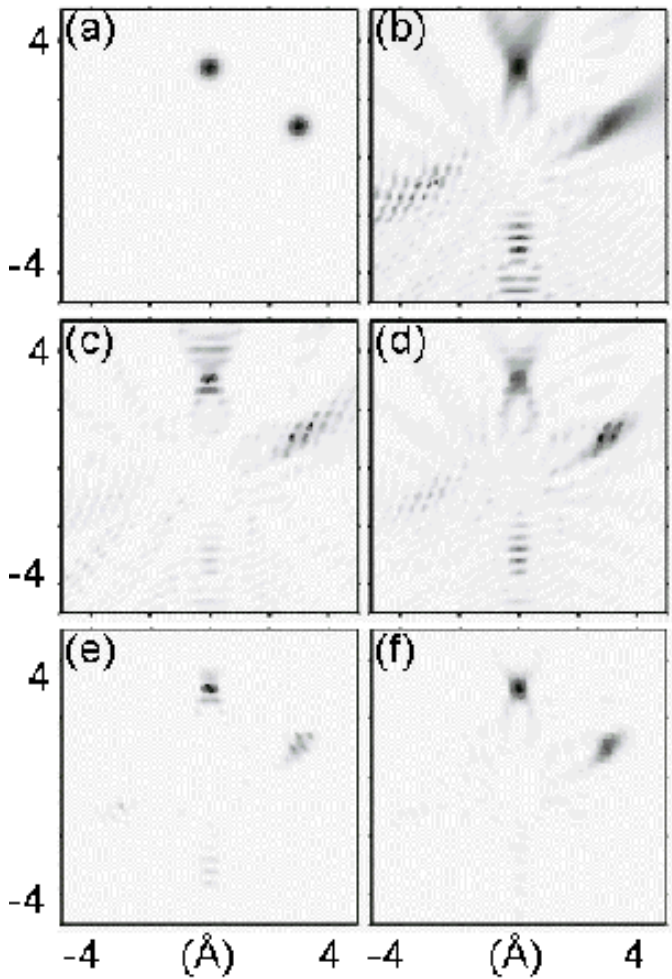}}
\caption{real space image of 
(a) the original charge density distribution - ${\rm{\bf F}}_0=50$, 
(b) Barton reconstruction - ${\rm{\bf F}}=400$,
(c) Maximum Likelihood - ${\rm{\bf F}}=550$, 
(d) Maximum Entropy and - ${\rm{\bf F}}=300$, 
(e) constrained ML - ${\rm{\bf F}}=60$,
(f) Constrained ML - ${\rm{\bf F}}=60$.
(b-e) were obtained from the hologram at 17 keV in Fig. (\ref{fig2}a) while
(f) was obtained from the simulated holograms at 17 and 18 keV 
shown in Fig. (\ref{fig2}a,b). The colormaps are scaled for each picture.}
\label{fig1}
\end{figure}
\begin{figure}[htbp]
\centerline{\includegraphics[width=3.5in]{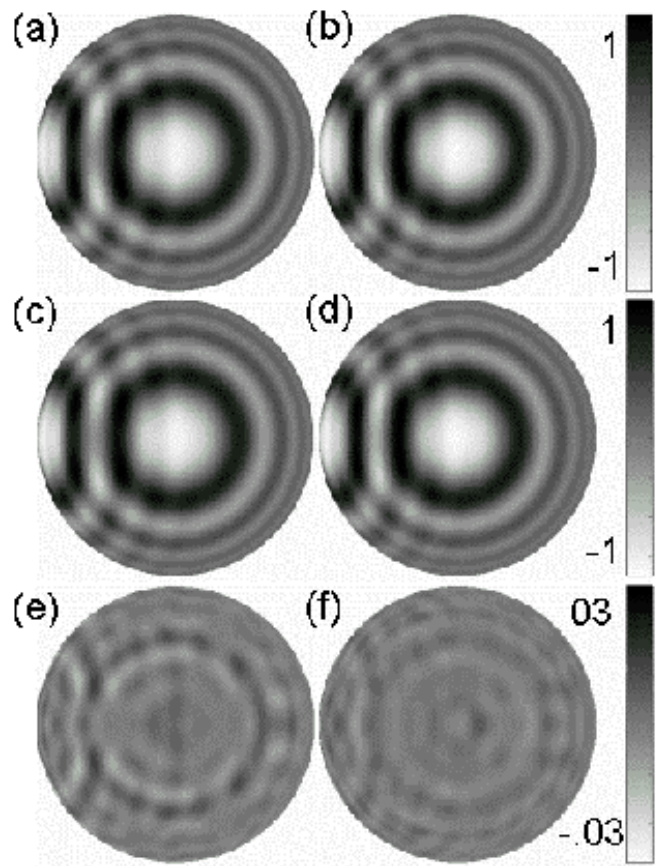}}
\caption{Top--Simulated holograms from two Gaussian charge distributions as shown in 
Fig. (\ref{fig1}a) at 17 keV and (\ref{fig1}b) 18 keV after smoothing. 
Center-- Holograms generated by the fitted charge density distribution at 
the same energies (\ref{fig1}c) 17 keV and (\ref{fig1}d) 18 keV. Bottom-- Difference between top 
and center: \ref{fig1}e = \ref{fig1}a - \ref{fig1}c and \ref{fig1}f =
\ref{fig1}b - \ref{fig1}d.}
\label{fig2}
\end{figure}
\begin{figure}[htbp]
\centerline{\includegraphics[width=3.5in]{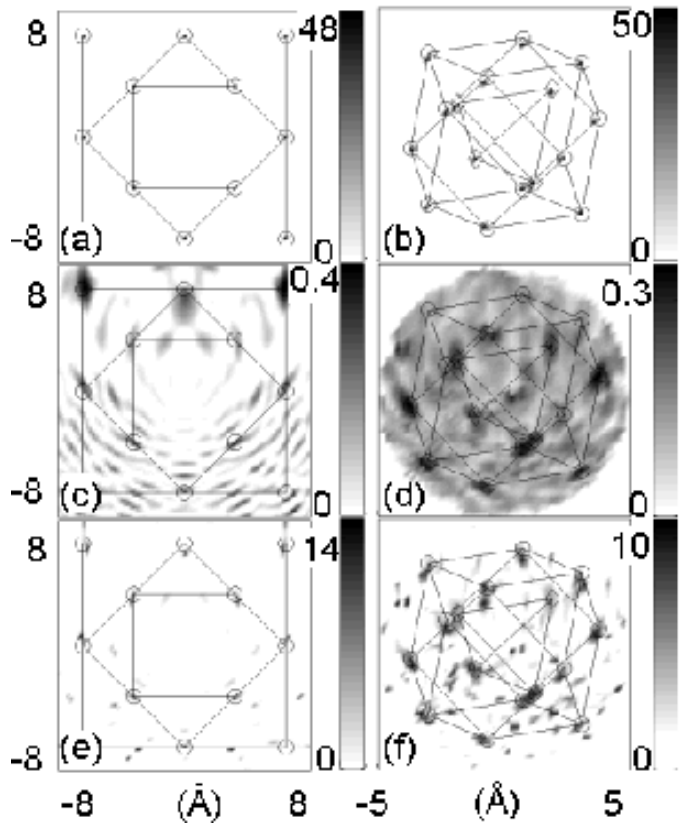}}
\caption{
Top- Original charge distribution (${\rm{\bf F}}_0=1.8\times \;10^4$) - 
(a) XZ plane, and (b) 3D view. 
Center- Standard holographic reconstruction (${\rm{\bf F}}=7\times \;10^4$), 
(c) and (d) the same views as (a) and (b), respectively. 
Bottom--Deconvolved charge density distribution 
(${\rm {\bf F}} =2.9\times \;10^4$), (e) and 
(f) show the same views as (a) and (b). The colorbar shows the maximum and 
minimum value. The 3D images are obtained by maximum voxel (volume-pixels) 
projection: the 3D matrix is rotated with respect to a plane (screen). On 
every pixel of the screen we project the maximum value of the voxels on top 
of the pixel itself. }
\label{fig3}
\end{figure}
\begin{figure}[htbp]
\centerline{\includegraphics[width=3.5in]{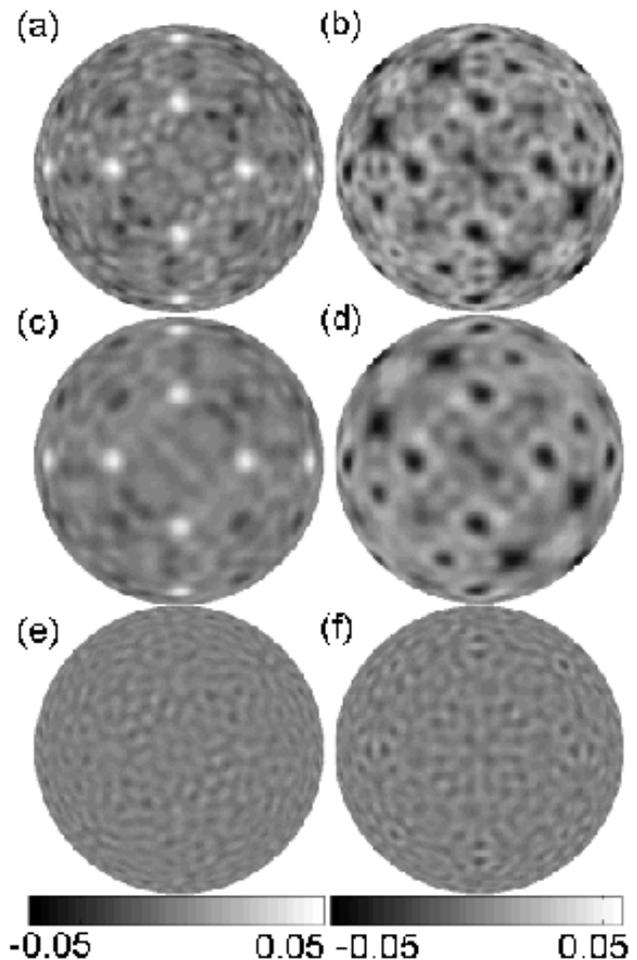}}
\caption{Top--Simulated holograms from a cluster of 1000 atoms at (a) 8 keV and (b) 9 
keV after smoothing. Center-- Holograms generated by the fitted charge 
density distribution at the same energies (a) 8 keV and (b) 9 keV. Bottom-- 
Difference between top and center: e = a - c and f = b - d}
\label{fig4}
\end{figure}
\begin{figure}[htbp]
\centerline{\includegraphics[width=3.5in]{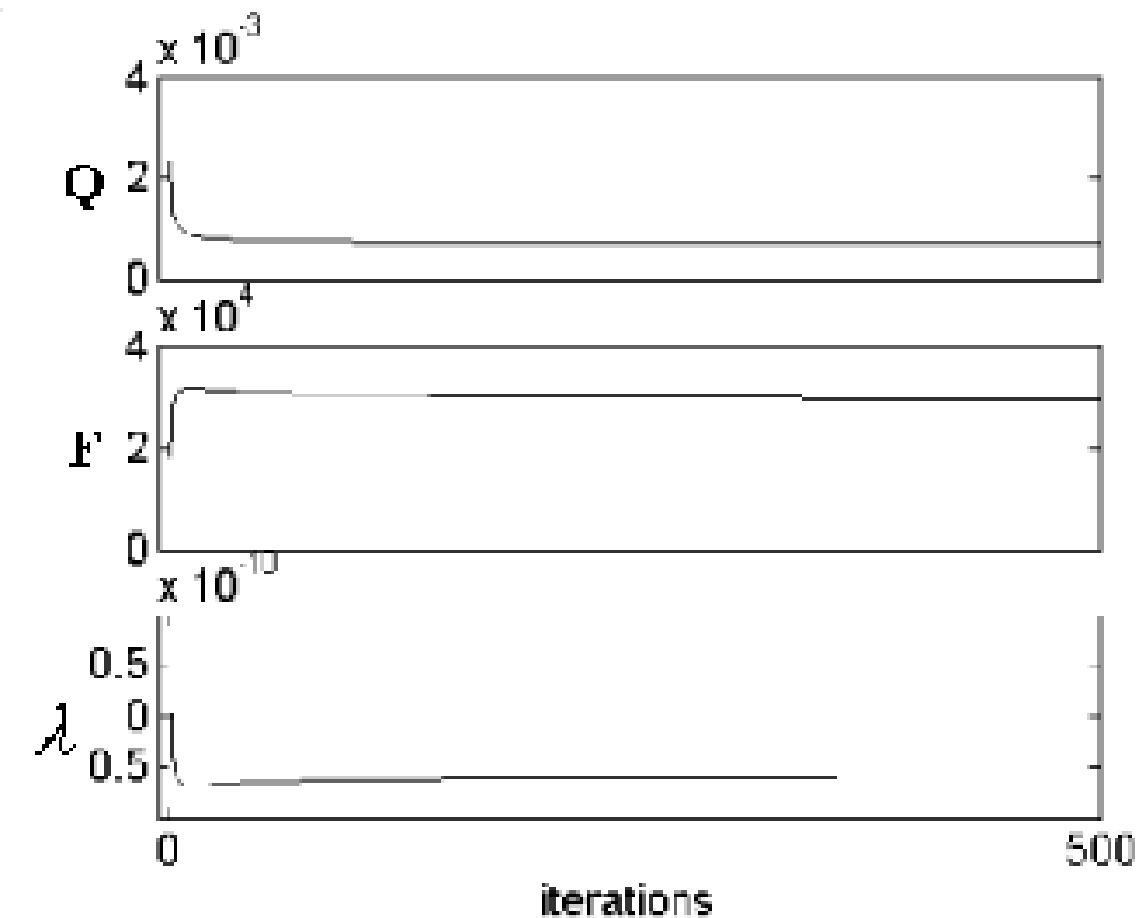}}
\caption{behavior of the fit quality (${\rm {\bf Q}}$) normalized to the volume in k-space, the number of electrons (${\rm {\bf F}}$) and the Lagrange 
multiplier $\lambda $. The fit is obtained quickly, while ${\rm {\bf F}}$ 
decreases much more slowly. The multiplier follows the behavior of ${\rm 
{\bf F}}$ as it tries to bring it back to ${\rm {\bf F}}_{0} \simeq 10^4$. }
\label{fig5}
\end{figure}
We have initially tried all the above mentioned deconvolution methods on a 
test system consisting of a charge distribution described by a 3D matrix of 
$50^3$ points with 0.2 {\AA} spacing. This matrix was set to 0 except for 
two Gaussian charge distributions centered in the x-z plane representing two 
atoms separated by 3.5 {\AA} as shown in 
Fig. (\ref{fig1}a). Figures 
(\ref{fig1}b-d) now show reconstructed images 
via several different methods obtained from a hologram 
(Fig. \ref{fig2}a) defined at an energy of 17 keV 
with the points distributed over a solid angle with azimuth varying from $0^0$ 
to $360^0$ and polar angle from $0^0$ (parallel to the z-axis) to $80^0$. The reconstruction in Fig. (\ref{fig1}f) 
has been obtained using both holograms at 17 keV and 18 keV 
(Fig. \ref{fig2}b). The standard direct Barton 
method in Fig. (\ref{fig1}b) shows a weak twin 
image, and some distortion of the imaged charge distributions, with the 
difference between Fig. (\ref{fig1}a) and 
Fig. (\ref{fig1}b) essentially being the PSF. The 
number of electrons estimated from the Barton method is one order of 
magnitude bigger than the original one. This could be ascribed to the fact 
that many of these electrons are not contributing to the hologram, as they 
cancel each other. In trials with the Van Cittert and ML 
(Fig. \ref{fig1}c) methods using the double pixel 
approximation, we have found that the convergence is very good: less then 10 
iterations are necessary for convergence. This is due to the fact that the 
holographic reconstruction is a relatively good image of the charge density. 
However the number of electrons used to fit the data was one order of 
magnitude bigger than that of the original image. Another relevant feature 
of this method is the fact that the image becomes more structured. This 
is due to the divergence in the inverse of the Hessian using single energy 
holograms when ${\rm {\bf H}}_{{\rm {\bf r}},{\rm {\bf r}}} = {\rm {\bf 
H}}_{{\rm {\bf r}}, - {\rm {\bf r}}} $.

Tests using the Lucy-Richardson method have shown bad convergence; further 
studies of this method need to be done, but one of the reasons could be some 
divergence caused by the denominator in equation (\ref{eq6}).

At this point we turned to the maximum entropy method. As we had already 
developed the ML algorithm with single and double pixel approximations, we 
could easily apply the algorithm developed by Cornwell and Evans 
\cite{Cornwell:1985} so as to employ the maximum entropy criterion. This 
algorithm tries to maximize the entropy ${\rm {\bf S}}$ while minimizing the 
likelihood ${\rm {\bf Q}}$ with a fixed number of electrons ${\rm {\bf F}}$ 
using Lagrange multipliers. Several tests have shown that this algorithm (at 
least the one implemented by us) was not able to minimize ${\rm {\bf Q}}$, 
${\rm {\bf S}}$ and ${\rm {\bf F}}$ at the same time; the result is shown in 
Fig. (\ref{fig1}d). However, we did find that the 
maximum likelihood and the minimum number of electrons produces a good 
quality reconstruction, with somewhat improved resolution, and a reasonably 
good ${\rm {\bf F}}$\cite{Marchesini:2001}; these results are shown in 
Fig. (\ref{fig1}e) and they can be compared with 
those in (\ref{fig1}b). The number of electrons is only 20{\%} bigger than the 
original one. 

The single energy iterative reconstruction shows how the divergence in the 
Hessian due to the twin image problem produces several artifacts, which are 
removed when using slightly different energies 
(Fig. \ref{fig1}f). The reason why the number of 
electrons does not correspond to the original ${\rm {\bf F}}_{0} $ even when 
applying the constraint is that after the steps are applied, we impose 
positivity by setting to 0 the negative values of the density. 

Finally, we applied the constrained ML method to the same case, using a 
double-energy data set. This result is shown in 
Fig. (\ref{fig1}f) and represents the most 
significant enhancement in image quality relative to the standard method.

To explore the constrained ML algorithm further, we carried out simulations 
based on a much larger and more realistic cluster of atoms representing 
CdTe; the charge density distribution in the neighborhood of the origin is 
shown in Fig. (\ref{fig3}a,b). We tried to keep a 
minimum amount of information to show the robustness of the method: We 
simulated the hologram produced by a CdTe cluster of 1000 atoms at only two 
energies of 8 and 9 keV. At these two energies the holographic 
reconstruction using the Barton method shows a poor image quality as shown 
in Fig. (\ref{fig3}c,d). We also introduced a randomly 
distributed noise level of $10^{ - 3}$ on every data point (corresponding to 
10$^{6}$ counts in each point), with the points distributed over a solid 
angle with azimuth varying from 0 to 360 degrees and polar angle from 0 
(perpendicular to the [001] surface normal direction) to 80 degrees, with a 
1-degree step size in each direction. Since we are interested in imaging the 
first neighbors, we finally carried out all imaging with a smoothed hologram 
based on 2 degree smoothing in every direction. Although the cluster size is 
several nanometers, we will only try to reconstruct up to 8 {\AA} setting to 
0 anything outside this range.

The holograms generated by this charge after smoothing are shown in 
Fig. (\ref{fig4}a) at 8 keV and Fig. (\ref{fig4}b) at 9 keV. We again choose as a first reference step and set of images 
the standard holographic reconstruction in 
Fig. (\ref{fig3}a,b). Comparison between the standard 
reconstruction and the deconvolved images are shown in 
Fig. (\ref{fig3}c,f).

The number of electrons within the reconstructed region is ${\rm {\bf F}}_0 = 
1.8\times \;10^4$ in the original picture, ${\rm {\bf F}} = 7\;\times 10^4$ in the standard reconstruction, 
and ${\rm {\bf F}} = 2.9\times \;10^4$ in the deconvolved image. Thus, the 
deconvolved image is a factor of 1.5 larger than the original image in number 
of electrons, but a factor of 2  better than the standard reconstruction. 
This result is worse than what we observed for the two Gaussian charges, and 
can be ascribed to the difficulty to fit point like structures (atoms) at 
the diffraction limit scale instead of smooth distributions. The peak 
heights of the charge density in the three images varies somewhat 
differently, the maximum number of electrons in one voxel in the original 
image is 50, 0.38 in the standard reconstruction, and 14 in the deconvolved 
image, as shown more quantitatively in the colorbars of 
Fig. (\ref{fig3}). The reason for this is that in 
the original image, every electron of one atom is concentrated in a single 
voxel, while the standard reconstruction has many electrons spread over many 
more voxels. The deconvolved image has a total of fewer electrons than the 
standard image, but much higher peaks, due to the enhanced resolution 
relative to the standard reconstruction. Comparing image quality between the 
standard (Fig. \ref{fig3}c,d) and 
deconvolved (Fig. \ref{fig3}e,f) 
approaches, we see a definite enhancement of resolution in the deconvolved 
images, even though some artifacts remain that are of comparable strength to 
the actual charge-density images. The holograms generated by the cluster in 
Fig. (\ref{fig4}a,b) are well described in the low frequency 
range by the fitted holograms in Fig. (\ref{fig4}c) (8 keV) 
and Fig. (\ref{fig4}d). Since in the iterations we only 
fitted the charge density within the first 8 {\AA} setting the distribution 
to 0 outside this range, only the low frequency components where fitted, as 
the residual difference shows in Fig. (\ref{fig4}e-f).

In Fig. (\ref{fig5}), we present some results 
dealing with the rapidity and type of convergence observed. The maximum 
likelihood is obtained within the first 10-20 iterations 
(Fig. \ref{fig5}-top). The total number of electrons is slowly reduced in the 
following $\sim 500$ iterations (Fig. \ref{fig5}-center). The Lagrange multiplier $\lambda $ follows the behavior of the 
constraint: as ${\rm {\bf F}}$ changes from the desired level ${\rm {\bf 
F}}_{0} $, $\lambda $ varies in order to bring ${\rm {\bf F}}$ back to the 
desired level.

\section{Conclusions and future possibilities}

In conclusion, we have demonstrated that iterative deconvolution can be 
applied to XFH imaging, thus improving the quality of the images beyond the 
diffraction limit, and that several levels of approximation can be used. 
From our test cases, the constrained maximum likelihood method appears to be 
the most promising. In these first theoretical simulations related to image 
improvement, we have approximated the scattering factor as isotropic, thus 
neglecting the angular dependence of Thomson scattering and near-field 
effects. As a computationally relevant issue, we have also approximated the 
Hessian to be a sparse matrix where only the two diagonals are non-zero. All 
of these approximations could be relaxed in future applications, including 
terms close to the diagonals, with of course additional complexity then 
being incurred in the deconvolution algorithm and the length of time it 
takes. For our calculations, the storage and calculation of the full Hessian 
of $80^6$ points required considerable memory, but computers capable of 
such calculations should be available as desktops in a few years. The 
gradient was also calculated neglecting the angular dependence of the 
scattering factor. Using these approximations, every iteration involved the 
calculation of a hologram from a matrix of $80^3$ voxels, the reconstruction 
and required about 5 minutes on a Pentium II computer, for a total of one 
night of calculation. Our image deconvolution has been obtained by enforcing 
a real and positive charge density distribution and limited number of 
electrons. However, looking ahead to further image improvement, once the 
image is deconvolved, one could make use of `atomicity' in the next step by 
placing atoms in the location of the peaks, and then performing an 
optimization using the full angular dependence of the scattering factor, 
while simultaneously adjusting the atomic number and position of these 
atoms. This study thus suggests several fruitful directions for further 
exploration and exploitation of image deconvolution in x-ray fluorescence 
holography that should considerably enhance its power for structural 
studies.

\begin{acknowledgments}
We would like to acknowledge A. Sz\"oke for 
fruitful discussions. This work was supported by the Laboratory Directed 
Research and Development Program of Lawrence Berkeley National Laboratory 
under the Department of Energy Contract No. DE-AC03-76SF00098. NATO 
Collaborative Linkage Grant for travel support.
\end{acknowledgments}

\appendix
\section{Appendix--Calculation of the Hessian and the gradient}
\label{sec:appendix}

The Hessian does not need to be known in great detail. It can be 
pre-calculated before starting the iterative algorithm neglecting the 
angular dependence of Thompson scattering factor $\frac{f_{T} }{r_{e} } = 
\frac{1 + \cos ^2\Theta _{{\rm {\bf k}},{\rm {\bf r}}} }{2}$ and using the following 
approximation $\int\limits_{\sigma \left( k \right)} {\cos ^2\left( {kr - 
{\rm {\bf k}} \cdot {\rm {\bf r}}} \right)\,k^2d^2{\rm {\bf \hat {k}}}} 
\simeq \textstyle{1 \over 2}\Omega _{k} $ and $\int\limits_{\sigma \left( k 
\right)} {\sin ^2\left( {{\rm {\bf k}} \cdot {\rm {\bf r}}} 
\right)\,k^2d^2{\rm {\bf \hat {k}}}} \simeq \textstyle{1 \over 2}\Omega _{k} 
$, with $\Omega _{k} = \int_{\sigma \left( k \right)} {k^2d^2{\rm {\bf \hat 
{k}}}} $, ${\rm {\bf \hat {k}}}$ is a unit vector, and $\sigma (k)$ is the 
region where the hologram is measured at a given energy. This yields:
\begin{eqnarray}
\nonumber
 {\rm {\bf H}}_{{\rm {\bf r}},{\rm {\bf r}}} &=& \sum\nolimits_{k} {\Delta 
k\int_{\sigma (k)} {\textstyle{{r^2} \over {r_{e}^{2} }}\cos ^2\left( {kr - 
{\rm {\bf k}} \cdot {\rm {\bf r}}} \right)\,k^2d^2{\rm {\bf \hat {k}}}} } , 
\\ 
\nonumber
 &\simeq& \textstyle{{r^2} \over {r_{e}^{2} }}\textstyle{1 \over 
2}\sum\nolimits_{k} {\Omega _{k} \,\Delta k} \,; \\ 
\nonumber
 {\rm {\bf H}}_{{\rm {\bf r}}, - {\rm {\bf r}}} &=& \sum\nolimits_{k} {\Delta 
k\int_{\sigma (k)} {\textstyle{r \over {r_{e} }}\cos \left( {kr - {\rm {\bf 
k}} \cdot {\rm {\bf r}}} \right)} } \\ 
\nonumber
 &&\times \cos \left( {kr + {\rm {\bf k}} \cdot {\rm {\bf r}}} 
\right)\,k^2d^2\hat {k} \\ 
\nonumber
 &=& \textstyle{{r^2} \over {r_{e}^{2} }}\sum\nolimits_{k} {\Delta 
k\int_{\sigma (k)} {\left[ {\cos ^2\left( {kr} \right) - \sin ^2\left( {{\rm 
{\bf k}} \cdot {\rm {\bf r}}} \right)} \right]k^2d^2{\rm {\bf \hat {k}}}} } 
, \\ 
\nonumber
 &\simeq& \textstyle{{r^2} \over {r_{e}^{2} }}\sum\nolimits_{k} {\left[ {\cos 
^2\left( {kr} \right) - \textstyle{1 \over 2}} \right]\Omega _{k} \,\Delta 
k} \\ 
 &\simeq& \textstyle{1 \over 2}\textstyle{{r^2} \over {r_{e}^{2} 
}}\sum\nolimits_{k} {\cos \left( {2kr} \right)\Omega _{k} \,\Delta k} \;;  
\label{eq16}
 \end{eqnarray}
The gradient is calculated by simulating the hologram $\chi ^{\left\{ n 
\right\}}\left( {\rm {\bf k}} \right)$ generated by a charge density 
distribution $\rho ^{\left\{ n \right\}}\left( {\rm {\bf r}} \right)$, and 
propagating back the hologram to the real space. Using $\bar {U}\left( {\rm 
{\bf r}} \right) = \left( {\textstyle{r \over {r_{e} }}} \right)^2U\left( 
{\rm {\bf r}} \right)$, eq. (\ref{eq8}) can be written as:
\begin{eqnarray}
\nonumber
 \nabla {\rm {\bf Q}} &=& 2\left[ {\bar {U}^{\left\{ n \right\}}\left( {\rm 
{\bf r}} \right) - \bar {U}\left( {\rm {\bf r}} \right)} \right]; \\ 
\nonumber
 \bar {U}^{\left\{ n \right\}}\left( {\rm {\bf r}} \right) &=& \eta ^\dag 
\left( {{\rm {\bf r}},{\rm {\bf k}}} \right) \cdot \chi ^{\left\{ n 
\right\}}\left( {\rm {\bf k}} \right) \\ 
\nonumber
 &\simeq& \frac{2r}{\Omega r_{e} }\Re\int_\sigma {e^{ - i\left( {kr - {\rm {\bf 
k}} \cdot {\rm {\bf r}}} \right)}\chi ^{\left\{ n \right\}}\left( {\rm {\bf 
k}} \right)d^3{\rm {\bf k}}} ; \\ 
\nonumber
 \chi ^{\left\{ n \right\}}\left( {\rm {\bf k}} \right) &=& \eta \left( {{\rm 
{\bf k}},{\rm {\bf r}}} \right) \cdot \rho ^{\left\{ n \right\}}\left( {\rm 
{\bf r}} \right) \\ 
 &\simeq& 2\Re\int {\textstyle{r \over {r_{e} }}e^{i\left( {kr - {\rm {\bf k}} 
\cdot {\rm {\bf r}}} \right)}\rho ^{\left\{ n \right\}}\left( {\rm {\bf r}} 
\right)d^3{\rm {\bf r}}}\, . 
\label{eq17}
\end{eqnarray}
These calculations can be very long as we need to sum over ${\rm {\bf 
r}}(x,y,z)$ for every ${\rm {\bf k}}\left( {k,\vartheta ,\varphi } \right)$. 
Supposing that the number of points in $x,y,z,\vartheta ,\varphi $ is $N 
\sim 100$ for every variable, we need to calculate the product in the 
integral of eq. (\ref{eq17}) for every possible value, {\it i.e.} $N^5$ products and 
sums. The computation of $\chi ^{\left\{ n \right\}}\left( {\rm {\bf k}} 
\right)$ can be performed faster if we separate the sum over the variables 
$x,y,z$.

We begin by calculating $\xi ^{\{n\}}\left( {\rm {\bf r}} \right) = 
\textstyle{r \over {r_{e} }}e^{ikr}\rho ^{\left\{ n \right\}}\left( {\rm 
{\bf r}} \right)\Delta r^3$; we can express Eq. (\ref{eq17}) as:
\begin{equation}
\label{eq18}
\chi _{n} \left( {\rm {\bf k}} \right) \simeq 2\Re\sum_{{\rm {\bf r}}} 
{e^{i{\rm {\bf k}} \cdot {\rm {\bf r}}}\xi ^{\{n\}}\left( {\rm {\bf r}} 
\right)} 
\end{equation}
\noindent
which is a simple Fourier transform. However the holograms are usually 
measured in spherical coordinates, and to obtain the values of $\chi _{n} 
\left( {\rm {\bf k}} \right)$ in the measured positions, we either need to 
perform an interpolation, or a convolution. Another approach is to use 
directly the spherical coordinates in the kernel:
\begin{eqnarray}
\nonumber
 e^{i{\rm {\bf k}} \cdot {\rm {\bf r}}} &=& e^{ikx\sin \vartheta \cos \varphi 
}e^{iky\sin \vartheta \sin \varphi }e^{ikz\cos \vartheta } \\ 
 &=& A_{k,x,\vartheta ,\varphi } B_{k,y,\vartheta ,\varphi } C_{k,z,\vartheta 
} 
\label{eq19}
 \end{eqnarray}
\noindent
and Eq. (\ref{eq18}) becomes:
\begin{eqnarray}
\nonumber
 \chi ^{\{n\}}\left( {k,\vartheta ,\varphi } \right) &\simeq& 
2\Re\sum\nolimits_{x} {A_{k,x,\vartheta ,\varphi } \sum\nolimits_{y} 
{B_{k,y,\vartheta ,\varphi } } } \\ 
&&\times \sum\nolimits_{z} {C_{k,z,\vartheta } \,\xi ^{\{n\}}\left( {x,y,z} 
\right)} \,.
\label{eq20}
\end{eqnarray}
Each sum requires $N^4$ calculation, reducing the time to perform the 
calculation by $\sim $N/3. The same trick can be applied for the holographic 
reconstruction.

\end{document}